# From populations to coherences and back again: a new insight about rotating dipoles

Sharly Fleischer[1,2*], Robert W. Field[1], Keith A. Nelson[1]

[1] *Department of Chemistry, Massachusetts Institute of Technology, Cambridge, Massachusetts 02139.*

[2*] *Department of Chemical Physics, Tel-Aviv University, Tel Aviv 69978, Israel. Email:sharlyf@post.tau.ac.il.*

Abstract:

The process in which light is absorbed by an ensemble of molecules obeys the fundamental law of conservation of energy - the energy removed from the light resides in the molecular degrees of freedom. In the process of coherent emission from excited molecules known as free-induction decay (FID), spectroscopic measurements of the emitted radiation are often conducted in order to gain insight into molecular structure and behavior. However, the direct influence of the FID emission on its molecular source is not measured directly. In this work we present experimental evidence from the 'molecular rotor perspective' of the consequences of terahertz-frequency FID emission from rotationally excited molecules. We show that when gas phase molecules transiently orient under field-free conditions, the energy radiated via FID is manifest as an abrupt reduction in excited rotational populations. The connection between coherent FID emission and stored energy leaves a particularly distinct signature in our measurements, but the results are generalizable throughout coherent spectroscopy and coherent control.

---------------------------

Laser assisted rotational control of gas phase molecules is an enabling component of 'molecular frame spectroscopy' in which measurements are made on ensembles of molecules prepared with their angular distribution maximized in a preferred direction of the laboratory frame. Most experimental efforts have gone toward creating molecular *alignment*, where internuclear axes lie preferentially parallel to a specified lab-frame *axis*. Control schemes based on superpositions of optical fields or on quasi-DC electric fields have induced net *orientation* of polar molecules, in which the permanent molecular dipoles point preferentially in the same laboratory *direction*. Weak single-cycle terahertz-frequency (THz) pulses, which are simultaneously resonant with multiple rotational transitions, have been used to exert short-duration torques on molecular dipoles. These result in transient net molecular orientation and an associated free-induction decay which has a temporal profile that consists of periodic spikes of THz-frequency emission [1, 2]. More intense single-cycle THz pulses induce not only net molecular orientation, which is a *first-order* effect proportional to the THz field amplitude, but also net molecular alignment and a non-thermal rotational population distribution, which are *second-order* effects, proportional to the square of the THz field amplitude [3]. By directly measuring both of these effects, we show that the amount of energy carried away by this periodic, coherent emission, is taken from the THz-induced non-thermal rotational populations. While our observations are facilitated by the unique spectral structure of molecular rotational transitions, the ideas presented here are fundamental to coherent spectroscopy and control. The results provide a 'molecule perspective' of resonant light-matter interaction, and illustrate the role of radiative coherences in the decay of excited populations, as dictated by energy conservation.

In a typical coherent spectroscopy experiment, an incident electromagnetic (EM) field imposes distinct phase relations between two or more eigenstates of an atom or molecule. These are known as coherences, and their presence is manifest in the time-dependent behavior of the system. In two-level systems regularly encountered in nuclear magnetic resonance [4] or optical spectroscopy [5], an electromagnetic (EM) field induces one-quantum coherences (1QCs) between two eigenstates. These 1QCs give rise to the emission of an oscillating EM field, known as free-induction decay (FID) at the two-level transition frequency, and with exponentially decaying amplitude. 1QCs can also develop from purely incoherent inverted populations, initiated by vacuum fluctuations and manifest as collective spontaneous emission, known as superradiance (SR) [6, 7]. Here we focus on a distinct phenomenon: emission from coherently excited polar molecular rotors subsequent to multi-level resonant excitation. Different from most FID-emission measurements, in which only the '*EM field component*' is measured, here we measure the '*matter component*' directly and observe the consequence of emission on the molecular ensemble from which it emanated. By monitoring the non-thermal rotational energy of the polar molecular ensemble throughout its evolution, we observe the energy uptake from the incident external field and subsequently, under external-field-free conditions, the depletion of this energy, released time-coincidentally with the emission of coherent radiation.

The system under study is a gas of carbonyl sulfide (OCS) molecules, for which many rotational transitions fall within the bandwidth of a strong single-cycle terahertz-frequency pulse [1,2,3] generated by optical rectification of a ~100 femtosecond duration optical pulse [8]. Figure 1b displays the spectrum of OCS rotational transitions at ambient temperature (black bars) and that of our typical THz pulse (green area). The temporal profile of the THz field, measured by the field-induced optical birefringence in an electro-optic crystal ("EO sampling" schematically shown in Fig. 1a) [9, 10] is shown in Fig. 1c. The OCS molecules are treated as rigid rotors with the Hamiltonian $\hat{H}_0 = \hat{J}^2/2I$, where $\hat{J}$ is the angular momentum operator and $I$ is the molecular moment of inertia, and have rotational transition frequencies given by $\omega_{J \to J+1} = 2Bc(J+1)$, where $B = h/8\pi^2 cI = 0.203$ cm$^{-1}$ is the molecular rotational constant, $c$ the speed of light in vacuum, and $J$ the rotational angular momentum quantum number. The THz field, $\vec{E}(t)$, acts on the molecular dipoles, $\vec{\mu}$, through the interaction potential, $\vec{H}_1 = -\vec{\mu}\cdot\vec{E}(t) = -\mu E(t)\cos\theta$, where $\theta$ is the angle between the dipole and the linear polarization axis of the field. The THz field induces coherences between adjacent rotational levels, $|J\rangle$ and $|J+1\rangle$ (one-quantum coherences or 1QCs), and these 1QCs start out in phase at time $t = 0$, i.e. during the THz excitation pulse. This initial preparation corresponds to an ensemble-averaged net orientation $\langle\langle\cos\theta\rangle\rangle > 0$ of the molecular dipoles. Due to their very different frequencies, the 1QCs dephase rapidly via their field-free evolution, and shortly after the end of the THz pulse, the initial net orientation is lost. However, since all of the rotational transition frequencies are integer multiples of the lowest transition frequency, $\omega_{0\to 1} = 2Bc$, (i.e. harmonics), the 1QCs rephase at time $T_{rev} = (2Bc)^{-1} = 82.2$ ps, known as the "revival" period. This sequence is repeated: with each revival, the rotational coherences rephase and a short-lived net orientation of the molecular dipoles manifest as a burst of coherent THz field emission [1,2,3]. This series of THz bursts is the FID, illustrated in Fig. 1c (for recent reviews on "quantum rotational revivals" see [11, 12] and references therein).

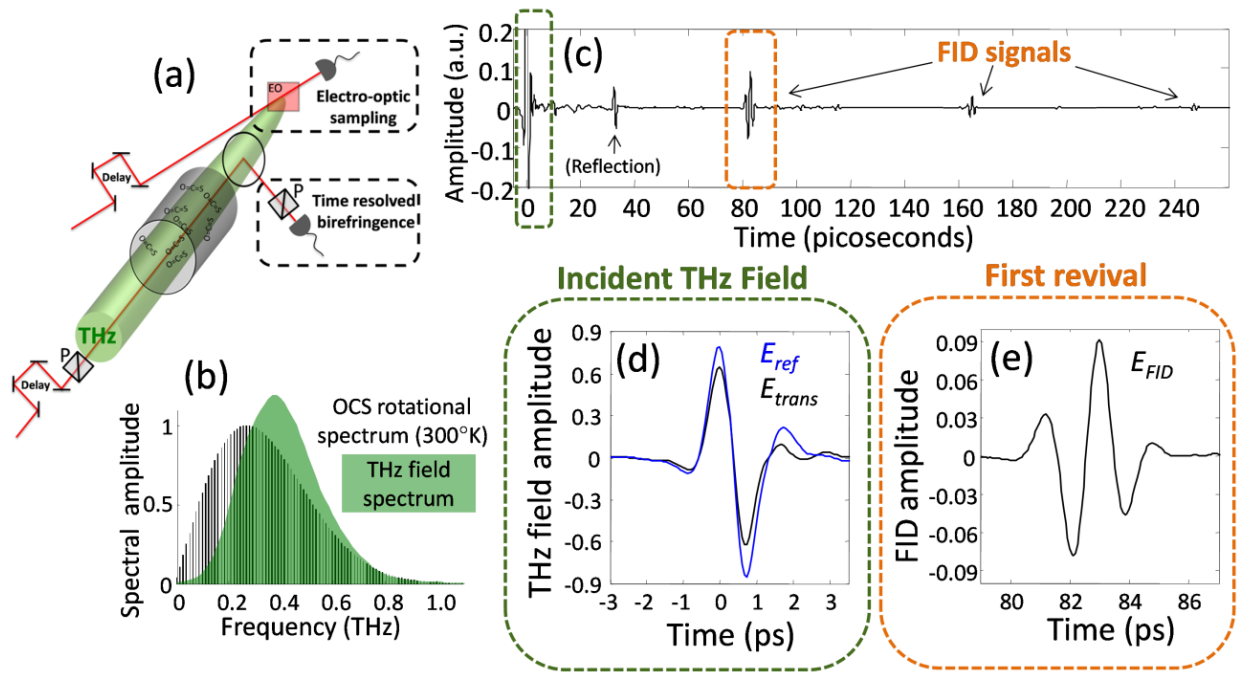

Figure 1: (a) Schematic experimental setup showing the THz pulse (large-diameter shaded beam) that drives molecular rotations, and the optical pulses (thin lines) used for electro-optic (EO) sampling measurement of the THz field time-dependent profile (including the incident pulse and the emitted FID bursts) and for time-resolved birefringence measurements. (b) Normalized rotational spectrum of OCS at 300K (black bars) and the frequency span of the single-cycle THz pulse (green area). (c) Experimental EO-sampling data of the output from the 380torr OCS gas cell. (d) Enlarged view of the incident THz field after its passage through the OCS cell (black curve) and through a reference vacuum cell (blue curve). (e) Enlarged view of the first THz FID burst, emitted at ~82ps following excitation.

From the EO-sampling data of Fig. 1c we calculate the fractional amount of energy emitted at the first FID emission burst, $E_{em} = \int_{80\,ps}^{86\,ps} |E_{FID}|^2\, dt$ ($E_{FID}$ in Fig.1e), relative to the energy absorbed from the incident THz field, $E_{abs} = \int_{-3\,ps}^{3\,ps} \left(|E_{ref}|^2 - |E_{trans}|^2\right) dt$, determined from THz transmission measurements through a 380-torr OCS gas sample ($E_{trans}$ in Fig. 1d) and through the same cell with the gas evacuated for reference ($E_{ref}$ in Fig. 1d). The fractional energy emitted in the first burst of FID radiation (first revival), $E_{em}/E_{abs} \sim 0.03$, is approximately 3% of the initially absorbed energy. The reductions in the field amplitude and radiated energy at subsequent revival times (FID signals at $t = 2T_{rev} = 164$ ps and $t = 3T_{rev} = 246$ ps) are mostly due to collisional decoherence processes in the 380 torr gas sample. Nevertheless, under decay- and decoherence-free conditions, the OCS is expected to radiate successively smaller amounts of energy since, with each and every emission event, the excess amount of energy stored in the molecular sample in the form of rotational energy must decrease.

In addition to the *electromagnetic-field perspective* described above, the system provides a unique signature of the molecule-field energy exchange from the *rotating molecule perspective*, in the form of molecular alignment measured via time-resolved optical birefringence (setup shown schematically in Fig. 1a). The energy absorbed from the THz field is stored as excess rotational energy that manifests as a non-thermal population distribution of rotational states, $|J,M\rangle$, where $M$ is the quantum number of the angular momentum projection onto the laboratory-frame $z$-axis. Rotational level population changes are induced by THz absorption, which are proportional to the incident THz pulse intensity, thus they express a *second-order effect*, namely, two field-molecule interactions. For molecules that are initially in eigenstates $|J,M\rangle$, the *first interaction* creates coherent superpositions of $|J,M\rangle$ and $|J+1,M\rangle$ states (1QCs) and the *second interaction* creates eigenstates, $|J+1,M\rangle$, i.e. populations, that are time-independent under interaction-free conditions (no external field and no collisions). The interaction potential induced by the linearly polarized THz field, $H_I(t) = -\mu E(t)\cos\theta$, does not change the quantum number $M$, and thus the resultant population distribution is non-thermal, not only in its THz-induced redistribution among energy levels ($J$ quantum numbers), but also because all of the excess population in the $J+1$ level is confined to the subset of $M$ states that also exist in the $J$ level, i.e. [$M=-J, -J+1,,,,, J-1, J$] but not [$M= -(J+1), J+1$] as shown in Fig. 2. In a thermal distribution (Fig. 2a), all of the degenerate $M$ states are equally populated which corresponds to an isotropic distribution of molecular axis directions, represented by a completely spherical probability distribution of the molecular axis directions (inset of Fig. 2a), and an ensemble-averaged alignment value of $\langle\langle \cos^2\theta \rangle\rangle = 1/3$. In the non-thermal distribution (Fig. 2b), there is a net alignment of the molecular axes along the THz polarization direction (the $z$-axis), represented by the non-spherical probability distribution which is elongated along the lab frame $z$-axis, and the alignment value of $\langle\langle \cos^2\theta \rangle\rangle > 1/3$. Classically, this results from the fact that the THz field exerts a torque that preferentially rotates each molecule in the plane defined by the molecule's original axis direction and the $z$-axis, thereby increasing the average projection of the molecular axis along the $z$-axis direction.

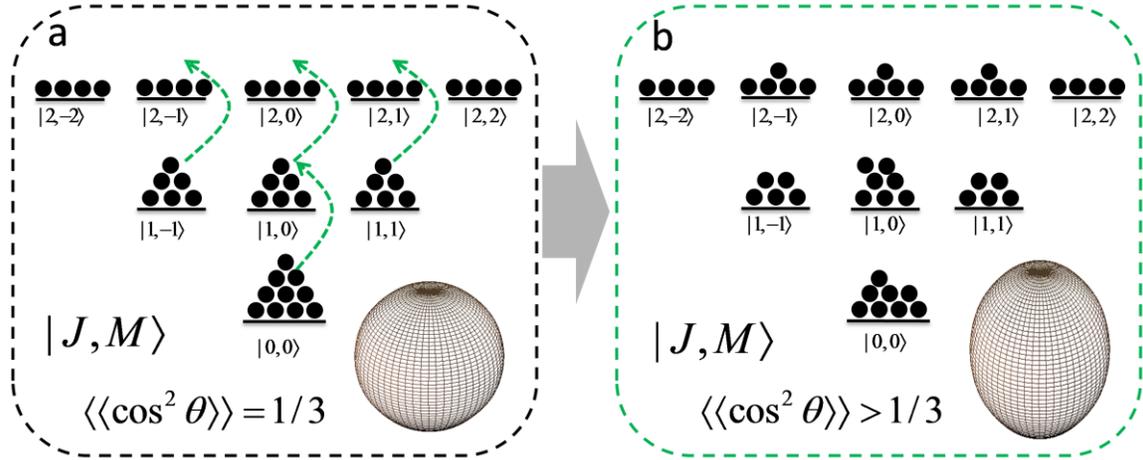

Figure 2: Qualitative presentation of the first three rotational levels, $J = 0$, 1, and 2, including the $M$ multiplicity. a) Thermal distribution - all $M$ states for each $J$ level are equally populated. b) Non-thermal distribution induced by a linearly polarized THz field. Populations are not transferred into states with the highest and lowest $M$ quantum numbers. The insets show the three-dimensional angular density for the thermal (spherically symmetric) and non-thermal (elongated along the z-axis) rotational state distributions.

This net alignment is manifest as optical birefringence, illustrated by Fig. 3a, recorded from OCS at a pressure of 150 torr. The time-dependent change in polarization of an 800-nm probe pulse, variably delayed with respect to the THz excitation pulse was measured as shown schematically in Fig. 1a. The slowly decaying signal is due to the non-thermal rotational population distribution. The sharp spikes in the signal are due to an additional effect that is caused by two instantaneous THz-molecule interactions that produce 2-quantum coherences (2QCs) between the $J$ and $J+2$ levels. All of the 2QCs start out in phase, yielding net birefringence at $t = 0$ (in addition to the contribution of the non-thermal rotational populations). The 2QCs rapidly dephase under field-free evolution and then rephase every half revival time, i.e. at $t = $ *(½, 1, 1½…)*$T_{rev}$. Two-quantum rotational coherences have been studied extensively [13, 14, 15] since they are produced through stimulated rotational Raman scattering by a short optical pulse, resulting in transient net molecular alignment but not orientation or the associated THz FID. The 2QCs are not the focus of the present work, but they are useful as markers for the fundamental revival times $T_{rev}$, $2T_{rev}$, etc. at which the FID bursts observed in our EO sampling measurements (fig. 1) are emitted. (The smaller signal spikes, preceding the main 2QC signals by ~ 5 ps, result from a weak pre-pulse, ~1% of the main THz pulse amplitude that arrives ~10 ps before the main pulse [16]). Signals like those in Fig. 3a, due to nonthermal population distributions, are expected to exhibit purely exponential decays [17, 18, 19] associated with the population decay rates of the excited rotational states. In our recently reported THz-induced optical birefringence measurements at OCS gas pressures above 350 torr [3], fast relaxation masked any more subtle features in the decay of THz-induced non-thermal populations and their corresponding birefringence signals.

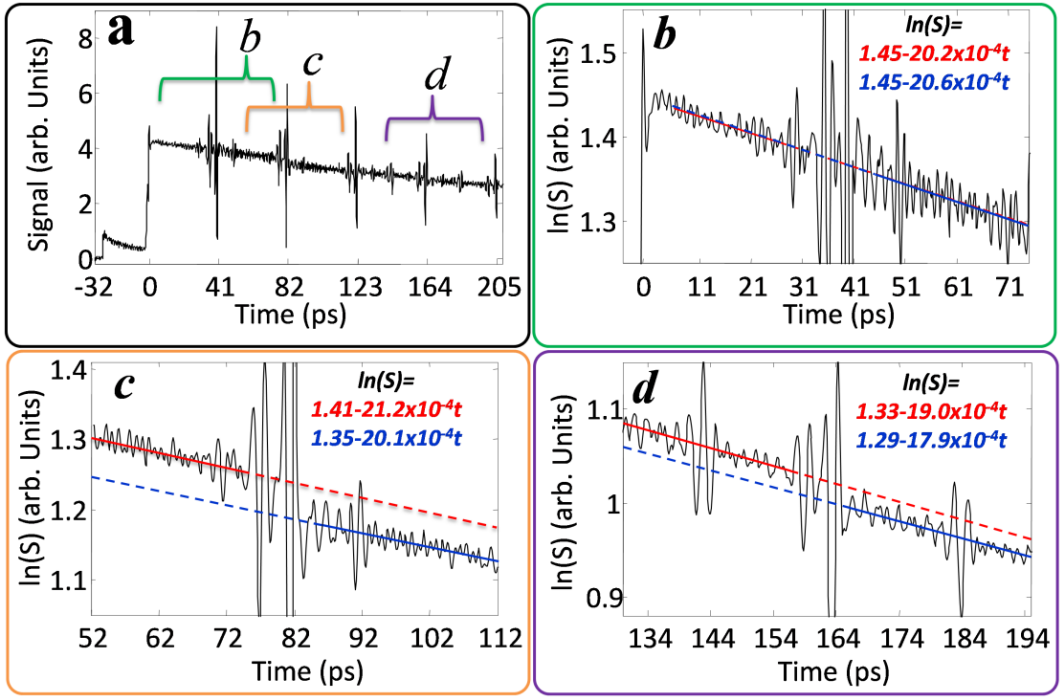

Figure 3: THz-induced birefringence in 150 torr OCS gas at room temperature. (a) Full time-domain scan, covering 2.5 revival periods (~200 ps). The steady-state signal level is due to THz-excited non-thermal rotational populations and the spikes in signal centered at times ½ $T_{rev}$ ~ 41ps, $1T_{rev}$ ~ 82 ps, etc. are due to two-quantum rotational coherences (2QC). (b-d) Semilog plots of THz-induced birefringence around selected times: (b) ½ $T_{rev}$, (c) $1T_{rev}$, (d) $2T_{rev}$. The linear fits (numerical values indicate the slopes in units of ps$^{-1}$ and the intercepts) compare the birefringence signal levels due to non-thermal rotational populations before (red lines) and after (blue lines) the selected times, with extrapolations shown as dashed lines. An abrupt decrease in signal level is apparent in (c) and (d).

Figures 3b-d show semilog plots of the data from Fig. 3a at the selected times of ½$T_{rev}$ = 41 ps, $T_{rev}$ = 82 ps, and $2T_{rev}$ = 164 ps following THz excitation (regions indicated by brackets). Each of the semilog plots was fit to straight lines based on the data collected in the regions before (red line) and after (blue line) the delay times of interest. The fit results are depicted as solid lines in the fitted regions and are extrapolated as dashed lines. The almost complete overlap between the red and blue lines throughout the time window of Fig. 3b confirms that no abrupt change in the non-thermal population occurred during this time window, and the decay in populations is purely exponential. This is expected since there is no FID emission at that time (see Fig. 1b at $t = 41$ ps). The same result was seen for the region around $\tfrac{3}{2} T_{rev}$ which is not shown here. However, in Figs. 3c and 3d, abrupt reductions in the non-thermal rotational population signals at times $t = T_{rev}$ and $2T_{rev}$ are evident. From the fits and the extrapolated signal levels at the times of FID emission, $t = T_{rev}$ and $2T_{rev}$, we deduced reductions in the non-thermal populations of ~ 5% and ~ 2%, respectively. The smaller decrease at $2T_{rev}$ is consistent with the reduced FID signal at that time (Fig. 1c) due to collision-induced dephasing of the coherences, which is evident in the data of Fig. 3a, where the 2QC signal amplitudes at $t = 82$ps and 164ps have a ~2:1 ratio.

Figure 4 depicts time-resolved birefringence measurements for three different OCS pressures (250, 150 and 65 torr). The results show that the percent decrease in the population signal at the time of FID emission ($\Delta$) is not related to the gas pressure, i.e. is independent of optical thickness. This excludes any effects due to non-uniform excitation along the length of the cell or THz propagation effects that may influence the magnitude of the abrupt decrease at higher pressures.

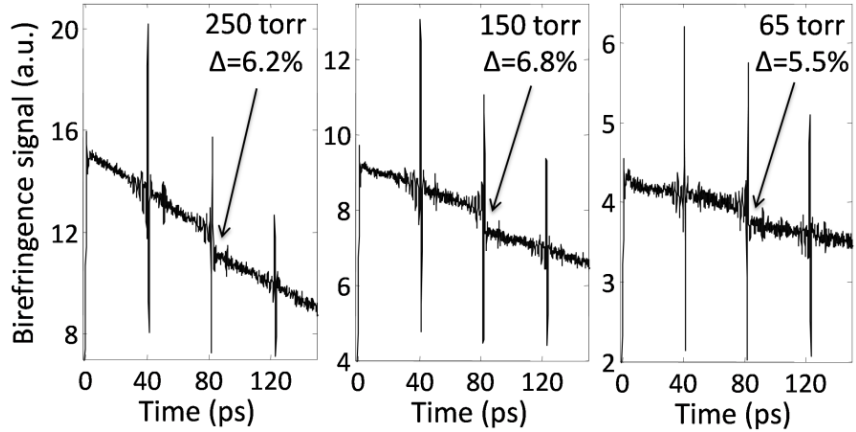

Fig.4: THz-induced birefringence scans in 250, 150, and 65 torr OCS gas samples. The decreases in the population signals are 6.2%, 6.8%, and 5.5%, respectively, i.e. all agree within the ~1% uncertainty of our measurement.

In the remaining of this paper, we analyze theoretically the interplay between coherent FID *radiation* and incoherent rotational *populations*. We introduce the radiative potential term, $\hat{H}_R(t) = \vec{\mu} \cdot d\vec{P}(t)/dt = \vec{\mu} \cdot \vec{\varepsilon} \times d\langle\langle \cos\theta \rangle\rangle/dt$, where $\vec{P}$ is the net polarization of the sample under external-field-free conditions, i.e. the orientation of the molecular dipoles, and $\vec{\varepsilon}$ is a unit vector in the direction of orientation (+z direction, the THz-field polarization axis). $\hat{H}_R(t)$ describes the dipole interaction of the molecule with its own emitted field upon net orientation. Here, we used the rotational Hamiltonian $\hat{H}_{FID} = \hat{H}_0 + \hat{H}_1 + \hat{H}_R$, where $\hat{H} = \hat{H}_0 + \hat{H}_1 = \hat{J}^2/2I - \vec{\mu}\cdot\vec{E}(t)$ is the Hamiltonian used in previous works to simulate the rotational responses of polar molecules to an external EM field [3,16,20, 21, 22, 23, 24, 25]. $\hat{H}_{FID}$ allows us to describe not only the burst of FID emission that occurs each time there is net molecular orientation, but also the effect of this emission on the molecules themselves. This is in addition to the field-free ($\hat{H}_0$) and interaction ($\hat{H}_1$) terms in the typically used $\hat{H}$. We simulated the molecular rotational responses by iterative solution of the Liouville-Von Neumann equation, $\partial\rho/\partial t = -i/\hbar[H,\rho]$, for the rotational density matrix, $\rho$. For additional information about the model and simulation, see Supplementary Information, section #2.

Figure 5 depicts the results of two simulations performed with the typically used Hamiltonian, $\hat{H}$ (red curves), that lacks the radiative potential term, and with $\hat{H}_{FID}$ (blue curves) that includes it. A simplified rotational system, with rotational constant $B = 1$ cm$^{-1}$ and T = 70K, was used to span a range of thermally populated transition frequencies comparable to those of OCS at 300K, but with fewer rotational levels, in order to facilitate the graphical presentation of the effects discussed here. The parameters of the THz field used in the simulation are given in the caption of Fig. 5. In both the red and blue simulated responses, the THz field terminates at ~2 ps.

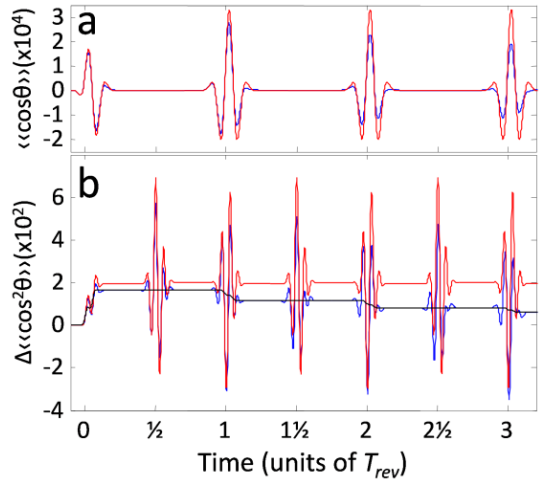

Figure 5: Simulated responses of a thermal ensemble of linear molecules with parameters $B = 1$ cm$^{-1}$ and $T = 70$K, excited by a single-cycle THz field with a 1.5 ps duration (full width at half max) Gaussian temporal envelope and a central carrier frequency of 0.5 THz. (a) Net molecular orientation as a function of time for $\hat{H}$ (red) and $\hat{H}_{FID}$ (blue). (b) THz-induced net molecular alignment $\Delta\langle\langle\cos^2\theta\rangle\rangle = \langle\langle\cos^2\theta\rangle\rangle - 1/3$ as a function of time for $\hat{H}$ (red) and $\hat{H}_{FID}$ (blue). The black curve depicts the contribution of THz-excited non-thermal rotational populations to the degree of alignment (without the 2QCs) and shows the abrupt decay in the excited populations with each FID emission.

The simulations were performed under collision-free conditions, i.e. without including collision-induced population relaxation or decoherence. With $\hat{H}$, successive orientation revivals are equal in amplitude, as shown in Fig. 5a. The modified Hamiltonian $\hat{H}_{FID}$ results in successive reductions of the orientation amplitudes for every revival period. The simulated change in the degree of alignment $\Delta\langle\langle\cos^2\theta\rangle\rangle$ is shown in Fig. 5b. $\hat{H}$ yields no attenuation in either the population or 2QC contributions to alignment (the phases of alternate 2QC signals are switched by $\pi$, as noted previously [12]). With $\hat{H}_{FID}$, the 2QC signal (blue curve) is attenuated slightly with time. Most significantly, *the population signal (black curve, overlapping the steady-state contribution to the blue curve) shows an abrupt attenuation with each successive revival period, signifying the decay of rotational populations due to energy exchange between the molecules and the emitted THz field.* The abrupt attenuations occur only at integer revival times, synchronized with the net orientation of the molecular ensemble and the corresponding FID emission events (Figs. 5a and 1b). Simulations performed with $\hat{H}_{FID}$ for OCS at 300K and with our experimental THz field parameters predict a 7.5% decrease in the population signal due to FID emission at $t=1T_{rev}$ (under decoherence-free conditions). At the finite pressures of the experiment, the molecules experience collision-induced decay and decoherence, resulting in exponentially decreasing population signal and FID emission amplitudes that yield a slightly smaller attenuation of the population signal, ~6%, in good agreement with the experimental data shown in Figs. 3c and 4.

*Generality of the present results*
We have presented experimental evidence for the interplay between radiative rotational coherences and excited rotational populations from the *rotating molecule's perspective*. We have shown that the energy lost via free-induction decay radiation is balanced by a decrease in non-thermal rotational populations. This result, obviously consistent with conservation of energy, only becomes apparent through direct experimental measurement of the populations' contributions to signals, which second-order in the incident field, and their FID-induced depletion, which is also second-order in the incident field. In our experiments, the harmonic structure of the rotational transition spectrum of linear molecules, and the short duration of the incident THz pulse, combine to constrain the FID to short-duration emission events that occur with specific periodicity. This permits a clear distinction between the radiation-induced decay of populations, revealed in a particularly graphic fashion as discrete, step-wise decay events, and other population relaxation mechanisms (mostly due to intermolecular collisions), which result in purely exponential decays, devoid of any abrupt features. Most FID measurements, including those from vibrational, electronic and nuclear spin coherences, do not show abrupt temporal features that would yield clear signatures in time-dependent population measurements. However, the underlying phenomenon presented here is general and can be expected to occur in myriad cases where radiative coherences are induced along with excited populations, i.e. in wide-ranging cases of coherent resonant excitation of a system. The significance of the general effects discussed here for any particular system will depend on the relative time scales for 1QC dephasing (through which the FID emission comes to an end) and population relaxation. The effects will be most significant in systems in which nonradiative relaxation rates are significantly lower than radiative relaxation rates.

The effect described here is collective among all of the rotational states, i.e. the energy removed via emission from each state is dependent on all of the coherently excited states, and hence all of the 1QCs and rotational states are coupled through the emitted field as described by the radiative potential $\hat{H}_R(t)$. In this respect, the effect is reminiscent to superradiance, which also depletes incoherent populations and couples the emitting species via the coherently radiated field. However, unlike superradiance following incoherent excitation, which emerges at a time that cannot be predicted precisely and which is phase-coherent but with a phase that cannot be anticipated, in the case of coherent excitation, the temporal and phase profiles are dictated by the free-induction-decay and thus, predictable. More importantly, there is no threshold that must be surpassed in order for the radiative effects on incoherent populations to occur, since the energy that is carried away by FID radiation is directly balanced by a reduction in excited rotational populations, as dictated by conservation of energy. Practical considerations such as low excitation field levels, where only first-order coherences (i.e. FIDs) may be observable above noise or the inability to directly monitor the excited population levels may obscure the observation of coherent radiative depletion of excited populations. However, as soon as second-order effects become observable, the measured population decays may include significant contributions due to coherent FID emission. In most cases, especially in two level systems where the FID emission is associated with only one frequency component, these effects will not show temporal features as distinct as those of multi-level rotational systems discussed here, but that should not obscure their generality or their role in maintaining energy conservation.

Acknowledgments:
We wish to acknowledge Dr. Harold Y. Hwang, Jian Lu, Yan Zhou and David Grimes (MIT), Dr. Assaf Tal (Weizmann Institute), Dr. Ronny Costi, and Dr. Einat Tirosh (Tel-Aviv University) for stimulating discussions. This work was supported in part by the Office of Naval Research Grant No. N00014-09-1-1103 (K. A. N.) and by the National Science Foundation Grant No. CHE-1111557 (K. A. N.) and CHE-1058709 (R. W. F.).